\documentclass[conference]{IEEEtran}
\usepackage{latexsym}
\usepackage{graphicx}
\usepackage{amsmath}
\usepackage{amsfonts}
\usepackage{amssymb}
\usepackage{textcomp}
\usepackage{float} 

\begin{document}

\title{A coherence study on EEG and EMG signals}

\author{
Giulia Cisotto, Umberto Michieli, Leonardo Badia \\
Dept.\ of Information Engineering, University of Padova \\
via Gradenigo 6B, 35131 Padova, Italy \\
email: { \{giulia.cisotto, leonardo.badia\}@dei.unipd.it }
}

\maketitle

\begin{abstract}
The aim of this study is to investigate bursts-related EEG signals in a focal hand dystonia patient.
Despite of considering time domain and frequency domain techniques as mutually exclusive analysis, in this contribution we have taken advantage from both of them: particularly, in the frequency domain, coherence was used to identify the most likely frequency bands of interaction between brain and muscles; then, in the time domain, cross-correlation was exploited to verify the physiological reliability of such a relationship in terms of signal transmission delay from the centre to the periphery.
Our preliminary results suggest - in line with recent literature - that activity in the high $\beta$ band (around 30 Hz) could represent an electroencephalographic correlate for the pathological electromyographic bursts affecting the focal hand dystonia condition. Even though a future study on a larger sample is needed to statistically support these preliminary findings, this contribution allows to think of new kinds of rehabilitation from focal hand dystonia that could target the actual electroencephalographic correlate of the pathology, i.e. phenotypically expressed by bursts, with the consequence of a relevant functional improvement.
\end{abstract}

\IEEEpeerreviewmaketitle

\section{Introduction}

There is a wide literature robustly reporting on how a voluntary motor output is prepared and driven by the central nervous system, the sensorimotor circuit, in particular. Brain signals recorded at several depths of the brain, i.e. (from outer to inner layer) the scalp, the dura, the cortex and the basal ganglia, have shown robust patterns of activation and deactivation in specific regions, frequency bands and time periods.

Specifically, power decrease at the contralateral hand-related side (to the movement) in the so-called $\mu$ and $\beta$ bands, around 10 Hz and 20 Hz, respectively, occurs as soon as 1s before movement onset. This phenomenon is known as \emph{event-related desynchronization} (ERD) \cite{Pfurtscheller1999}.

In the time domain, time-locking each electroencephalographic (EEG) signal on its corresponding electromyographic (EMG) activation onset, and averaging among several responses, a typical waveform could be observed: indeed, the so-called \emph{readiness potential} starts as soon as 1.5 to 1 s before movement onset with a slow decrease of signal amplitude; other known components follow, each of them with a specific clinical meaning. This complex behavior, overall, is labelled as \emph{movement-related cortical potential} (MRCP) \cite{Shibasaki1980}.

In case of neuro-motor pathologies, ERD and MRCP could become carriers of important information related to abnormal behaviours of the patient.
Particularly, in case of motor disorders, where movements are often involuntarily produced, a central, i.e. conscious, control has been suggested \cite{Lin2009} \cite{Ruiz2009} \cite{Shibasaki2005}, but not consistently proved and accepted, yet.

The \emph{Jerk-locked back averaging} (JLBA) technique has been effectively employed on MRCP to identify the central origin of a specific kind of involuntary movements observed in myoclonus, Tourette's syndrome and other psychogenic motor disorders \cite{Shibasaki1975}.
Thanks to such technique, a sharp biphasic waveform could be consistently seen on the averaged EEG signal, especially at the central and contralateral areas. Moreover, this EEG \emph{potential} anticipated the EMG onset by 15 to 20 ms on average \cite{Shibasaki2012} \cite{Avanzini2016}.

Later on, other techniques have been utilized to quantify the influence of brain activity on the motor output. Moreover, coherence between EEG and EMG has been computed in many studies for different kinds of patients: widely-known as \emph{cortico-muscular coherence} (CMC) \cite{Mima2002}, it is usually evaluated during sustained contractions at a predetermined level, e.g. 15 to 20 percent of the maximal voluntary contraction, in order to ensure stable motor units engagement.

However, CMC was also employed in few recent works on Parkinson's disease resting tremor \cite{Timmermann2003}\cite{Hellwig2001}: in those studies, no movement was accomplished by the patients but their postural tremor was recorded by EMG together with synchronous EEG or magnetoencephalographic (MEG) activity.

A clear peak of activity at the frequency of tremor and its second harmonic, around 5 Hz and 10 Hz respectively, could be seen in the EMG power spectrum; moreover, CMC showed a statistically significant peak of coherence between the EEG signal recorded from the contralateral hand-related scalp area and the fingers extensor muscle.

Therefore, in case of rhythmic pathological behaviour, a relationship between central and peripheral activity could be significantly quantified at rest, too.

The aim of this study is to apply a similar concept to the investigation of bursts-related EEG signals in one focal hand dystonia (FHD) patient.

FHD is a movement disorder that causes people who are affected by it to experience an abnormal and involuntary co-contraction of the agonist and antagonist muscles of the hand and the forearm. It has been shown to originate in the central nervous system and to cause abnormal patterns of brain activation, especially in the $\beta$ band (as suggested by recent literature) \cite{Tecchio2008}.

Bursts are abrupt and giant involuntary muscular contraction events that typically affect EMG of this kind of patients, especially at rest when they largely exceed the very low background activity.


In this contribution, CMC as well as cross-correlation have been computed between EEG and EMG to assess, both in the frequency and in the time domain, the effect of a central driver onto the motor output.
Our preliminary results show the influence of EEG on pathological EMG oscillatory activity. Coherence was employed to identify the most involved frequencies, while cross-correlation was used to support the physiological meaning of such EEG-EMG relationship.

In the rest of the paper, section II will present the methods of the study, section III the most interesting preliminary results, while the final section IV will discuss them in comparison with existing literature on the topic along with an overview of some limitations to be overcome in the future; a perspective view for clinical applications in motor rehabilitation will be provided, also.

\section{Materials and Methods}
One FHD patient and one healthy subject (HS) were involved in this pilot study.
The EEG was recorded from one monopolar EEG channel placed on C3, the standard location of the International 10-20 System over the left hemisphere corresponding to the brain region related to the functioning of the (dominant) right-hand.
The EMG was recorded from one bipolar channel placed on the \emph{abductor pollicis brevis}, the intrinsic hand muscle responsible for the abduction of the thumb.
Both signals were sampled with a sampling frequency of 1 kHz and quantized at 16 bit.
In the experiment, the participants were sitting quietly on a comfortable chair in front of a screen placed 1 meter apart from them, on a table. They were simply required to rest with opened eyes for about 3 minutes with their limbs laying on the table in front of them.


At a first glance, it was possible to assess a clear difference between the two EMG signals: in the FHD patient, the amplitude of the signal assumed values up to $\pm$ 200 \textmu V, while its power spectral density (PSD) took significant values in the frequency band $[5$, $200]$ Hz.
However, in case of HS, the amplitude of the EMG signal did not exceed $\pm$ 20 \textmu V, with a significant PSD extended from $5$ to $50$ Hz.

In the offline analysis, signals were preprocessed to limit their frequency range in the frequency band of interest. Specifically, the EEG was filtered through an elliptic filter of order $24$ with a passband of $[4,45]$ Hz. The EMG was processed by a high-pass elliptic filter of order $11$ with cut-off frequency at $5$ Hz. A series of notch filters of order $14$ were used with cut-off frequencies at $50$ Hz and subsequent harmonics up to $350$ Hz were put to reduce the effect of the mains.

Then, CMC as well as cross-correlation have been computed between the EEG (otherwise labelled as signal $x[m]$) and the EMG (otherwise labelled as signal $y[m]$) signals, in order to assess the quantitative relationship between them, both in the frequency domain and in the time domain.

\subsection{Frequency domain analysis: cortico-muscular coherence}
The coherence of two discrete-time signals $x[m]$ and $y[m]$, regarded as stochastic processes, is given by:
\begin{equation}
{\rm Coh}_{xy}(f) \triangleq \frac{\mathcal{P}_{xy}(f)}{\sqrt{\lvert \mathcal{P}_x(f) \rvert} \cdot \sqrt{ \lvert \mathcal{P}_y(f) \rvert}},
\end{equation}
where $\mathcal{P}_{x}(f)$ is the PSD of $x[m]$ and $\displaystyle \mathcal{P}_{xy}(f)=\frac{1}{n} \sum_{i=1}^{n} X_i(f)Y_i^*(f)$ is the cross-power spectral density (CPSD) between $x[m]$ and $y[m]$.

In order to provide a statistically significant result, a confidence level $CL$ of $95$ $\%$, i.e. a critical level of $\alpha=0.05$, was obtained from the following formula \cite{Mima1999}:
\begin{equation}
CL = 1-(1-\alpha)^{\frac{1}{N-1}},
\end{equation}
where $N$ is the number of signal segments used to estimate the coherence value.

The PSD, the CPSD and the coherence values were estimated via the Fast Fourier Transform (FFT)-based Welch's method: specifically, the signal length was set to $L = 1024$ samples ($1.024$ s) and the number of FFT points to $1024$ samples. Border effects were mitigated by a Hann sliding windowing with overlap of $50$ $\%$ \cite{Mitra2007}.

\subsection{Time domain analysis: cross-correlation function}
Generally speaking, given two discrete-time signals $x[m]$ and $y[m]$, their cross-correlation function is defined as:
\begin{equation}
{\tt r}_{xy}[n] \triangleq \sum_{m=-\infty}^{+\infty} x^*[m] y[n+m].
\end{equation}
Cross-correlation is particularly useful to evaluate the similarity between two signals as a function of the time shift $n$ (expressed in number of samples) of the second signal behind the first one.

In the present analysis, the absolute value of the correlation between the EEG and the EMG signals computed at its maximum and normalized by the square root of the product of the signals energies $E_x$ and $E_y$ was evaluated. Therefore, the quantity:
\begin{equation}
\displaystyle {\tt r}_{max}=\frac{\max (r_{xy}[n])}{\sqrt{E_x E_y}}.
\end{equation}
was considered as a measure of similarity between the two signals.

Moreover, the lag $n$ which the maximum was found at was taken into account as a measure of the transmission delay from the brain to the muscle, i.e. the time taken for a motor command to travel from its origin in the central nervous system to the target effector at the periphery.

Particularly, $71$ pairs of EEG and EMG signals were extracted from the whole EEG and EMG recordings of the FHD patient. They were selected empirically as examples of bursty EMG activity (with their corresponding EEG).
The duration of these signals was variable ($0.70 \pm 0.66$ s): all of them were included in the analysis to keep into account the variability of the burst events.

\begin{figure}[t]
\centering
\includegraphics[width=0.5\textwidth]{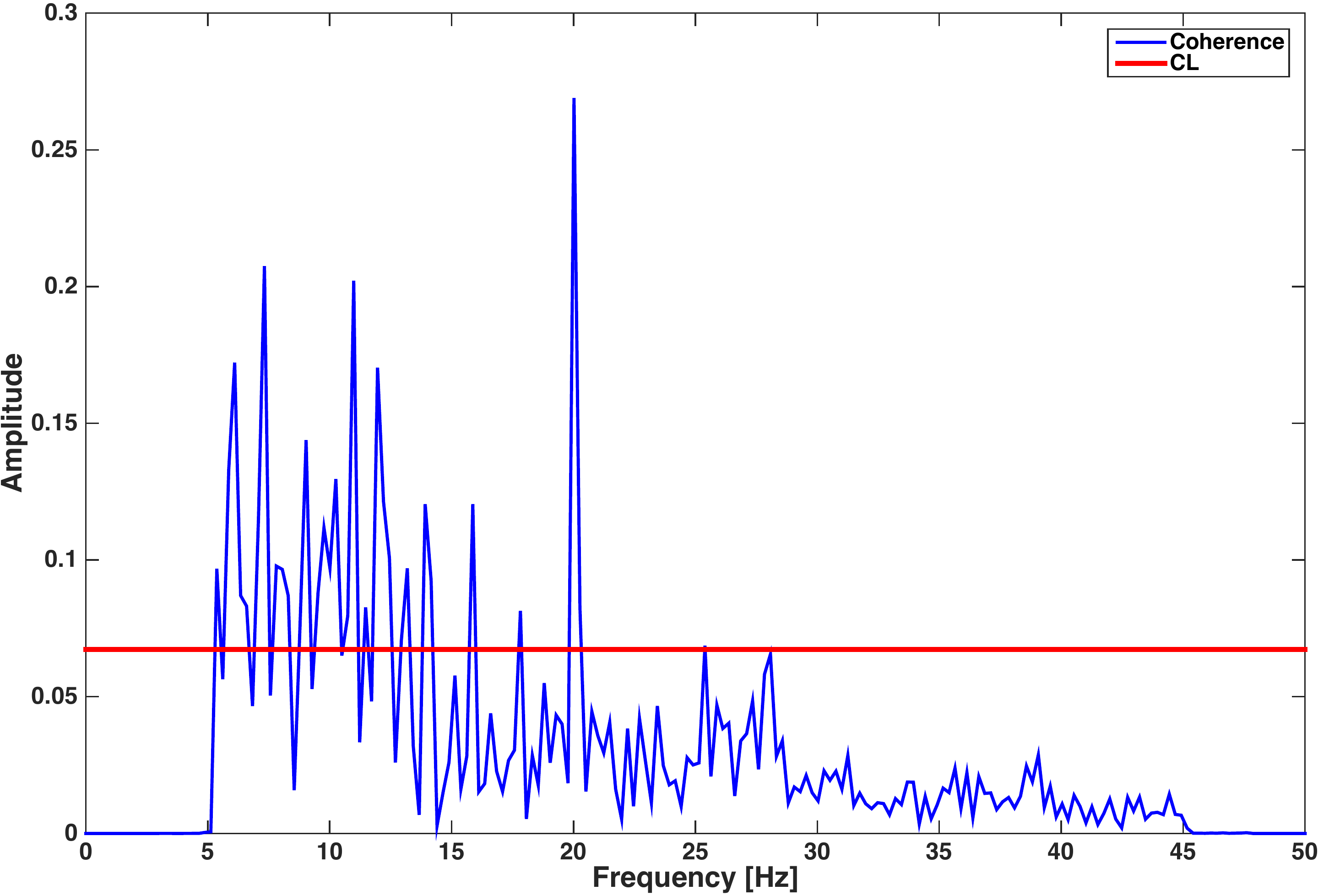}
\caption{Absolute value of CMC for the healthy participant ($CL=0.067$ with $N=44$ and $\alpha=0.05$).} \label{fig:coerenza_sano_totale}
\end{figure}

\begin{figure}[]
\centering
\includegraphics[width=0.5\textwidth]{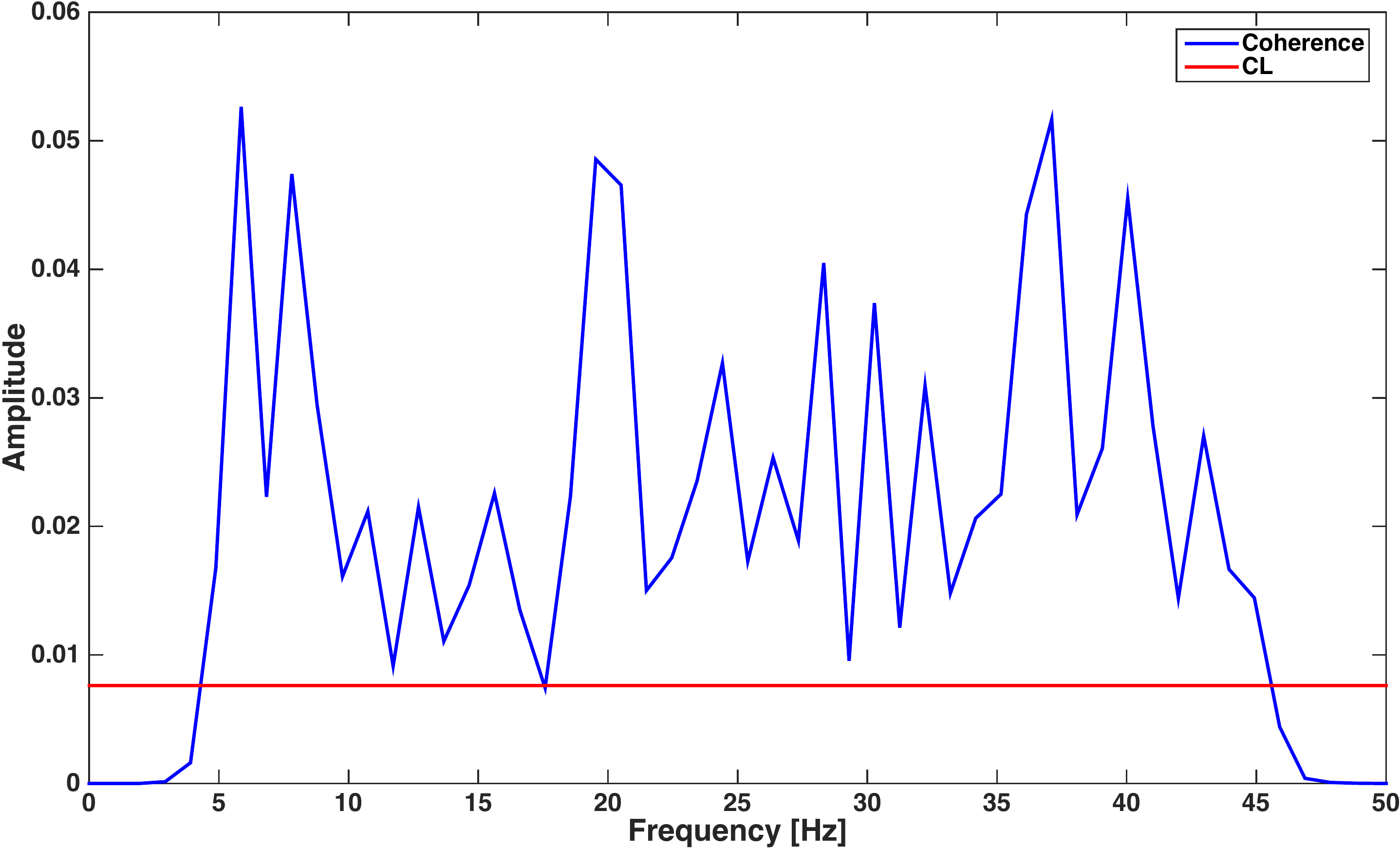}
\caption{Absolute value of CMC for the pathological subject ($CL=0.008$ with $N=393$ and $\alpha=0.05$).} \label{fig:coerenza_patologico_totale}
\end{figure}

To support the physiological meaning of the EEG-EMG coherent components, the cross-correlation function was computed between the narrow-band EEG signals filtered in the high $\beta$ band, i.e. between $26$ and $31$ Hz, and the EMG signal limited to $250$ Hz by a band-pass filter with frequency band $[5, 250]$ Hz.

\section{Results}

\subsection{EEG-EMG coherence}
In this section the results are shown in regards to the CMC measure for both the HS and the FHD patient.
In the case of the healthy participant, the CMC spectrum could be seen in \figurename~\ref{fig:coerenza_sano_totale}.
It has to be noted that peaks above the confidence level can be observed in the frequency range between $5$ and $20$ Hz, only, with a particularly strong coherence at $20$ Hz.

On the other hand, the CMC spectrum of the FHD patient is reported in \figurename~\ref{fig:coerenza_patologico_totale}.
It can be observed that a larger frequency band contribute to the coherence between EEG and EMG signals. It is also important to highlight the presence of peaks in the upper side of the spectrum, i.e. $[20, 45]$ Hz. This is probably due to the larger bandwidth of the pathological EMG of the patient, as mentioned in the previous section.

\begin{figure}[]
\centering
\includegraphics[width=0.5\textwidth]{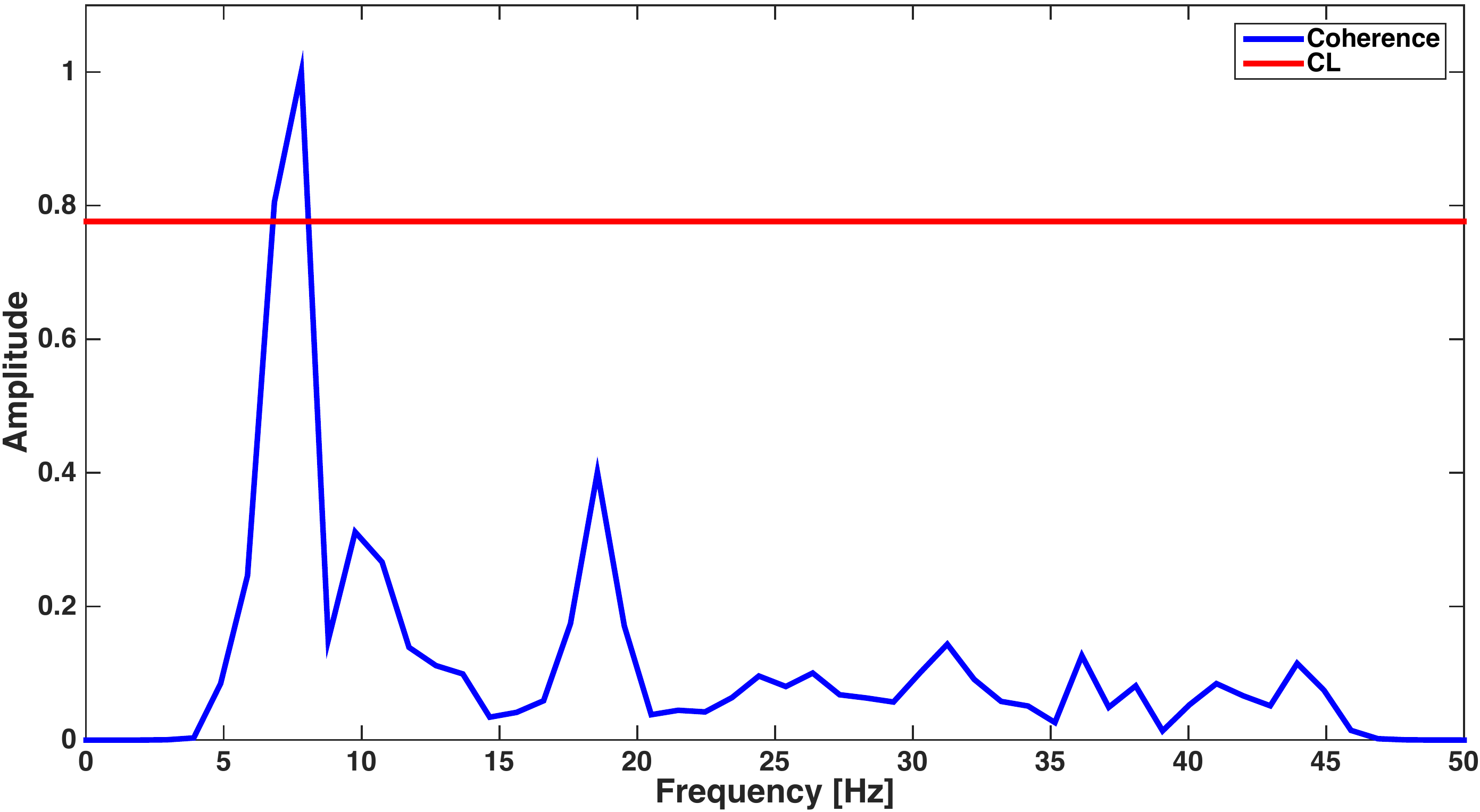}
\caption{CMC, in absolute value, between chunks of healthy-like EEG and EMG for the pathological subject ($CL=0.776$ with $N=3$ and $\alpha = 0.05$).}  \label{fig:125_coerenza}
\end{figure}

In order to confirm our hypothesis, we selected a portion of the whole recorded data where bursts mostly affected the EMG signal and evaluated the CMC in this particular case. As a further support, we selected another portion of EMG signal where healthy-like activity could be observed and computed CMC as well. 
Two typical examples of both situations are reported next.
\figurename~\ref{fig:125_coerenza} shows the coherence result when comparing two chunks of the EEG and EMG signals for the healthy-like case. Here, two main peaks can be seen at the frequencies of $8$ Hz and $18$ Hz, but no significant coherence values at frequencies higher than $30$ Hz.

\begin{figure}[]
\centering
\includegraphics[width=0.5\textwidth]{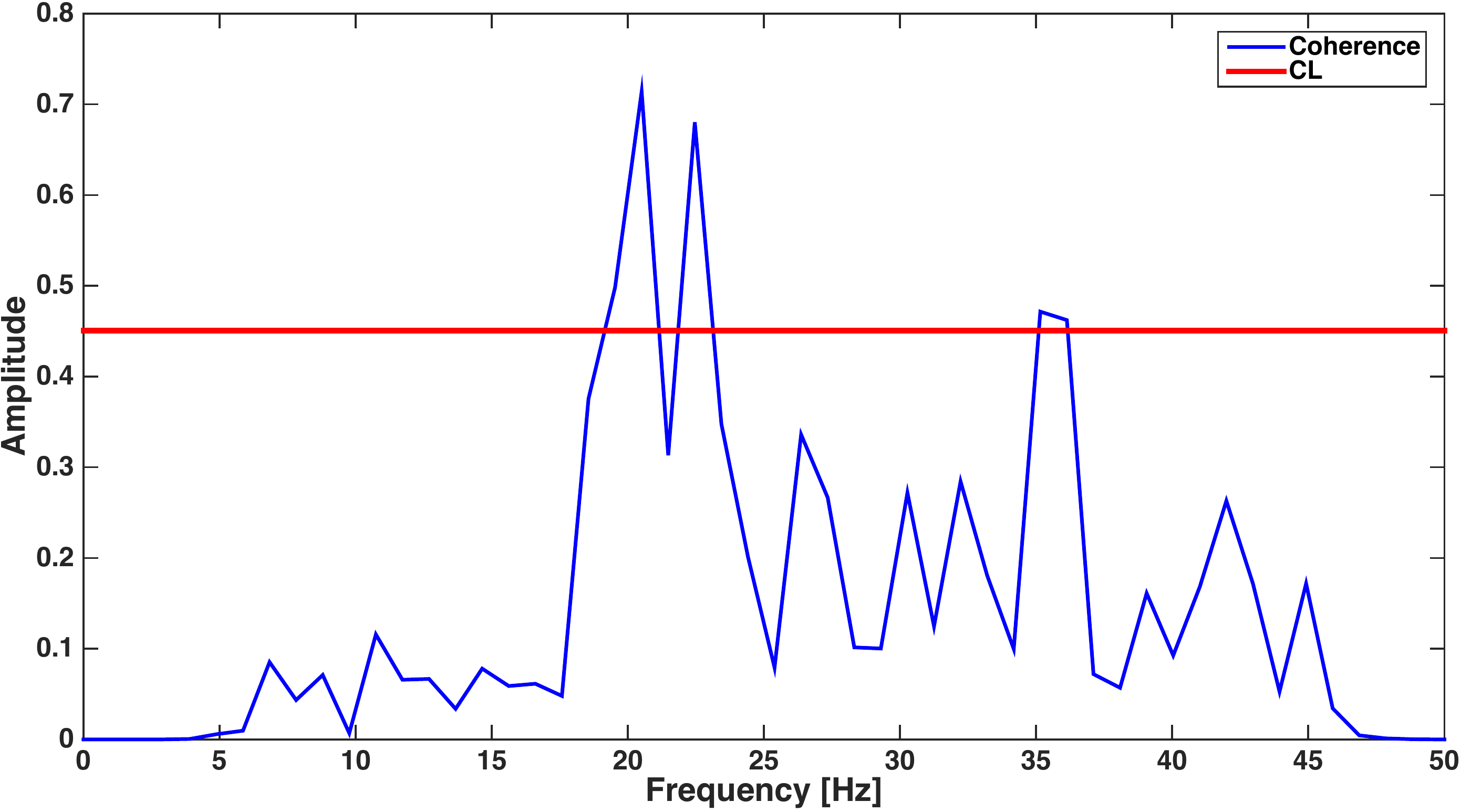}
\caption{CMC, in absolute value, between chunks of bursts-affected EEG and EMG for the pathological subject ($CL=0.451$ with $N=6$ and $\alpha = 0.05$).} \label{fig:117120_coerenza}
\end{figure}

On the contrary, \figurename~\ref{fig:117120_coerenza} reports the coherence spectrum in case of bursts-affected chunks. Significantly, the figure shows that coherence values at low frequencies are heavily reduced, whereas some peaks around $20$ and $35$ Hz appeared, hence  the hypothesis that higher frequencies components are related to bursty EMG activity in the FHD patient could actually be supported.

\subsection{EEG-EMG cross-correlation}
As mentioned above, cross-correlation of EEG and EMG was computed to investigate the physiological reliability on the relationship between the high $\beta$ band EEG component with the EMG.

\begin{figure}[t]
\centering
\includegraphics[width=0.5 \textwidth]{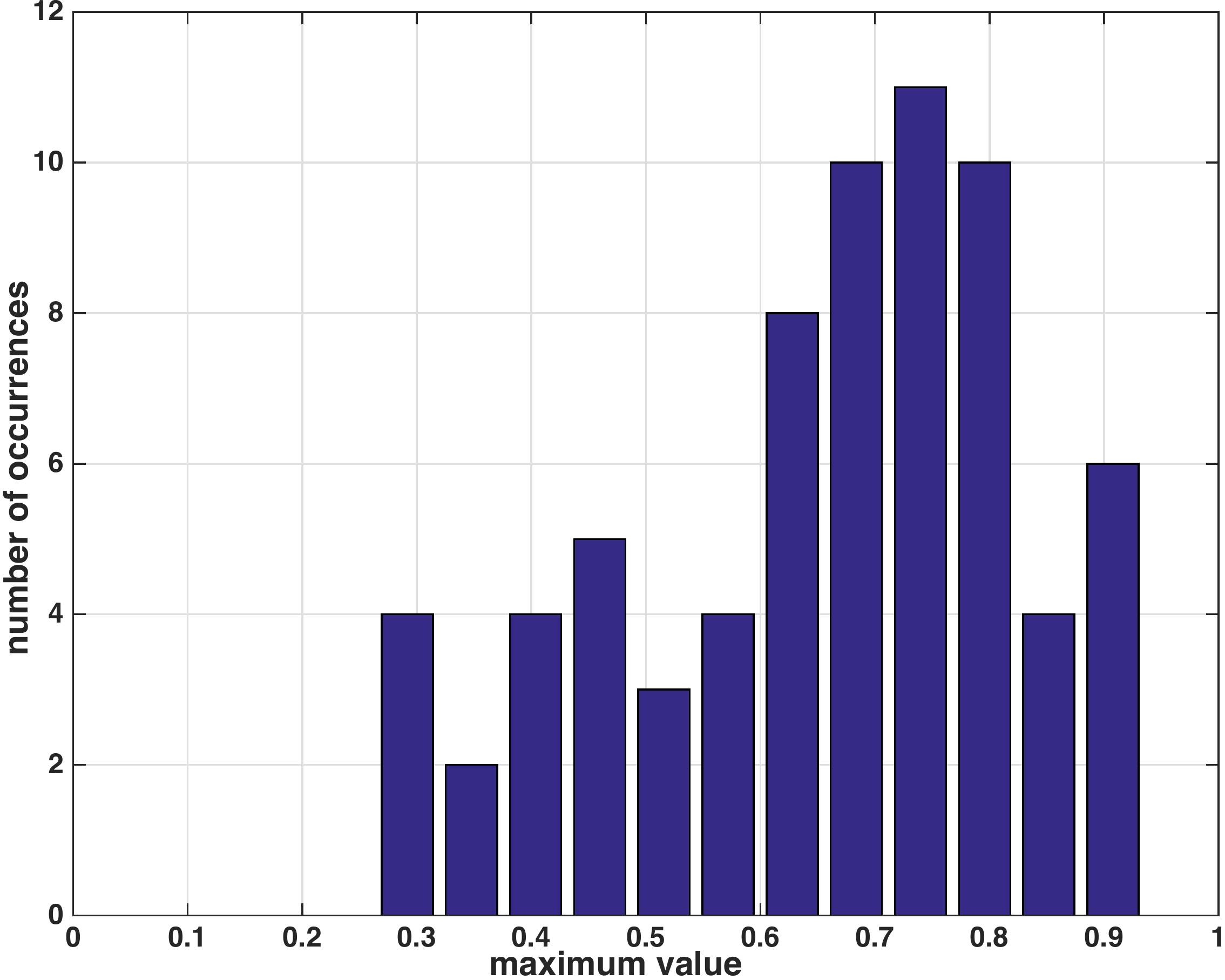}
\caption{Empirical distribution of the maximum value of the cross-correlation function of the $71$ pairs of EEG and EMG signals of the FHD patient.} \label{fig:isto_maxnorm}
\end{figure}

\figurename~\ref{fig:isto_maxnorm} reports the empirical distribution of the maximum values of the cross-correlation function found from the $71$ pairs of EEG and EMG signals.
Mean value was found of $0.683$, with variance of $0.0293$.
This result certainly shows a strong connection between the narrow-band EEG and the EMG, as indicated by the high average value.

\begin{figure}[t]
\centering
\includegraphics[width=0.5 \textwidth]{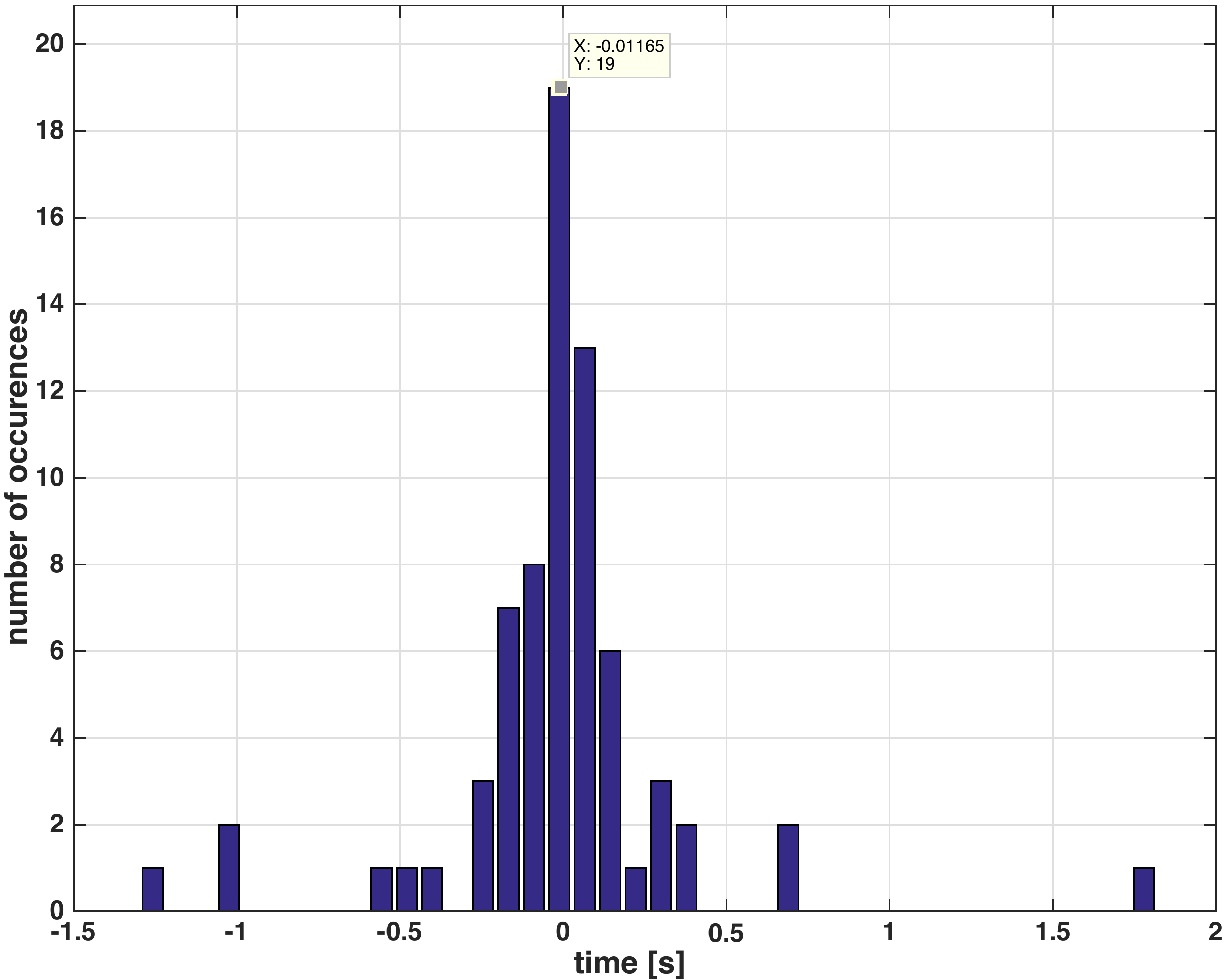}
\caption{Empirical distribution of the lags where the maximum of the cross-correlation was found.} \label{fig:isto_lag}
\end{figure}

Finally, \figurename~\ref{fig:isto_lag} displays the empirical distribution of the lag where the maximum value of the cross-correlation function of the $71$ pairs of EEG and EMG signals was found.
Mean value occurred at $-11.65$ ms.
As neural impulses propagate at a speed of about $100$ m/s, the transmission of signals from the brain to the  hand muscles could be estimated of about $10$ ms, which is in line with the results we achieved.

The standard deviation is considerably high (it was found to be about $100$ ms) but this could be explained because of the limited size of the data sample. Indeed, we expect that the tendency observed in this study could be further confirmed (with a reduced standard deviation), with an increased sample size.

\section{Discussion}

FHD and other movement disorders show jerks during movement or even during rest. Identifying possible central drivers of such abnormal muscular activity is a very relevant issue and could represent a key aspect for improving diagnosis and rehabilitation.
JLBA is the only technique that could directly reveal a causal relationship between an average brain activity and an average muscular response in many pathological cases.


However, JLBA is not always applicable and requires many chunks of EEG and EMG to be robust. Moreover, identification of jerks and averaging among jerk-locked EMG signals raise some technical questions in its application to FHD, as reported by other literature \cite{Brown1999}\cite{Xu2016}. Specifically, the following critical points were highlighted:
\begin{itemize}
\item high-frequency and irregularity of the jerks might prevent their correct identification (EMG activation should be absent for 100ms at least before the EMG burst);
\item similarly, the absence of giant somatosensory evoked components in many subjects could prevent success of the JLBA procedure;
\item frequency domain-based analysis have been successfully proposed to identify EEG correlates of myoclonic jerks, even in case where time domain JLBA has failed.
\end{itemize}

Since similarity could be noted between EMG bursts seen in FHD and jerks present in myoclonus, many considerations explained above hold true for analysis of EMG bursts in case of FHD.

Therefore, in this study we proposed the use of CMC in case of FHD, too. Besides, we do not believe time and frequency-domain analysis to be mutually exclusive but, on the contrary, we claim the opportunity to use them to complement each other.
Therefore, in this contribution, coherence analysis was used to identify the most likely frequency bands of central-periphery communications; then, cross-correlation function was employed to verify the physiological reliability of such relationship (as suggested by \cite{Shibasaki2012}).

In line with well-known literature about CMC \cite{Mima1999a} \cite{Mima1999}, we found a peak-component at 20 Hz ($\beta$ band) together with lower peaks at lower frequencies ($\theta$ and $\alpha$ bands) in the HS.

On the contrary, the FHD patient showed different CMC patterns in different behavioral scenarios: specifically, when a healthy-like recording period is selected, CMC spectrum displayed two major peaks, one around 5 Hz and the other one at 20 Hz. In this case, we could advance the hypothesis - supported by literature \cite{Timmermann2003} \cite{Hellwig2001} - that the slower component is related to the postural tremor affecting this subject, while the second one seems to reflect the healthy-like $\beta$ band synchronous activity between EEG and EMG that has been seen in the control subject, as well.
Worthy of notice is the fact that the tremor-related component significantly prevails onto the other one and it is the only one to exceed the confident level of $95$ $\%$.
When an EMG period heavily characterized by the presence of bursts is selected, we could observe two major frequency components in the CMC spectrum: the first one was around 20 Hz ($\beta$ band), it exceeded the confidence level of $95$ $\%$  and it could resemble the healthy-like $\beta$ band component of the normal condition. The second one, interestingly, was significant at the $95$ $\%$ as the first component, and could be suggested to represent the correlate of the bursty activity observed in the pathological EMG of the patient.

An average delay of $10$ ms of EMG behind EEG has been observed through the cross-correlation analysis and it is in line with physiology and other literature \cite{Mima1999}.

Overall, our preliminary results claimed that activity in the high $\beta$ band [25,35] Hz could represent an EEG correlate for the EMG bursts affecting the FHD condition.

Even though the small size of the data sample could be a limitation to the study, our conclusions were primarly driven by our data analysis, but their reliability was strongly supported by existing literature about similar studies accomplished with larger samples of data and patients, even though in slightly different diseases, e.g.myoclonus and Parkinson's disease.


Even though a future study on a larger sample is needed to statistically support these preliminary findings, this contribution allows to think of new kinds of rehabilitation interventions \cite{VonCarlowitzGhori2015} \cite{Hashimoto2014a} for focal hand dystonia patients that could target the actual EEG correlate of the pathology, i.e. phenotypically expressed by bursts, with the consequence of a relevant functional improvement.

\section{Conclusions}
The aim of this study is to investigate bursts-related EEG signals in a focal hand dystonia patient.
Despite of considering time domain and frequency domain techniques as mutually exclusive analysis, in this contribution we have taken advantage from both of them: particularly, in the frequency domain, cortico-muscular coherence was used to identify the most likely frequency bands of interaction between brain and muscles; then, in the time domain, cross-correlation was exploited to verify the physiological reliability of such a relationship in terms of signal transmission delay from the centre to the periphery.
The most interesting result suggested that the high $\beta$ band activity in the EEG could be responsible for the bursty activity observed in the EMG.
Even though a future study on a larger sample is needed to statistically support these preliminary findings, this contribution allows to think of new kinds of rehabilitation interventions for focal hand dystonia patients that could target the actual EEG correlate of the pathology with consequence improvement of the motor functions.

\bibliography{CMC}

\end{document}